\documentclass[draft]{agujournal2019}
\usepackage{url} 
\usepackage{lineno}
\usepackage{hyperref}
\usepackage{amsmath} 
\usepackage[inline]{trackchanges} 
\usepackage{soul}
\draftfalse

\journalname{Enter journal name here}

\begin{document}

%
%


\title{Improving Typhoon Predictions by Integrating Data-Driven Machine Learning Models with Physics Models Based on the Spectral Nudging and Data Assimilation}

%
%




\authors{Zeyi Niu\affil{1,2}, Wei Huang*\affil{1}, Lei Zhang\affil{1}, Lin Deng\affil{1}, Haibo Wang\affil{1}, Yuhua Yang\affil{1}, Dongliang Wang\affil{1}, and Hong Li\affil{1}}


\affiliation{1}{Shanghai Typhoon Institute, and Key Laboratory of Numerical Modeling for Tropical Cyclone of the China Meteorological Administration, Shanghai, China.}
\affiliation{2}{Department of Atmospheric and Oceanic Sciences and Institute of Atmospheric Sciences, Fudan University, Shanghai 200438, China.}




\correspondingauthor{Wei Huang}{hangw@typhoon.org.cn}



\begin{keypoints}
\item An ML-driven hybrid typhoon model combing the Pangu model and WRF using spectral nudging
\item The ML-driven hybrid typhoon model outperforms ECMWF IFS in typhoon intensity forecasts
\item Data assimilation can further improve the performance of the ML-driven hybrid typhoon model
\end{keypoints}

%
%

%
%


\begin{abstract}
With the rapid development of data-driven machine learning (ML) models in meteorology, typhoon track forecasts have become increasingly accurate. However, current ML models still face challenges, such as underestimating typhoon intensity and lacking interpretability. To address these issues, this study establishes an ML-driven hybrid typhoon model, where forecast fields from the Pangu-Weather model are used to constrain the large-scale forecasts of the Weather Research and Forecasting model based on the spectral nudging method (Pangu\_SP). The results show that forecasts from the Pangu-SP experiment obviously outperform those by using the Global Forecast System as the initial field (GFS\_INIT) and from the Integrated Forecasting System of the European Centre for Medium-Range Weather Forecasts (ECMWF IFS) for the track forecast of Typhoon Doksuri (2023). The predicted typhoon cloud patterns from Pangu\_SP are also more consistent with satellite observations. Additionally, the typhoon intensity forecasts from Pangu\_SP are notably more accurate than those from the ECMWF IFS, demonstrating that the hybrid model effectively leverages the strengths of both ML and physical models. Furthermore, this study is the first to explore the significance of data assimilation in ML-driven hybrid dynamical systems. The findings reveal that after assimilating water vapor channels from the Advanced Geostationary Radiation Imager onboard Fengyun-4B, the errors in typhoon intensity forecasts are reduced.

\end{abstract}

\section*{Plain Language Summary}
This study introduces a novel approach to predict typhoons by combining machine learning (ML) and traditional physics-based models. By integrating forecasts from the Pangu-Weather with the Weather Research and Forecasting (WRF) model using a technique called Spectral-Nudging, we significantly improved the accuracy of typhoon track and intensity predictions. Specifically, our method reduced the average track error for Typhoon Doksuri (2023) by 20.3\% compared to forecasts initialized using GFS analysis in four different initial times. The cloud patterns forecasted by Pangu\_SP align more closely with satellite imagery. Also, Pangu\_SP's predictions of typhoon intensity are notably better than those from ECMWF IFS and Pangu-Weather. Additionally, using water vapor data from the FY-4B AGRI satellite greatly improves the accuracy of typhoon intensity forecasts, underscoring the value of integrating data assimilation into this hybrid model.

%
%

%


%
%
%
%

\section{Introduction}
In recent decades, physics-based numerical weather prediction (NWP) models have achieved tremendous success in fields such as weather forecasting and climate change prediction (Kalnay, 2003; Kay et al., 2015). With the rapid development of techniques like model physical parameterization, data assimilation and ensemble forecasting (Hong et al., 2012; Geer and Bauer, 2011; Geer et al., 2018; Toth et al., 2003; Richardson et al., 2020), the performance of NWP models has been continuously improved. This gradual and incremental improvement through the accumulation of advanced techniques has been termed the “quiet revolution” of the NWP (Bauer et al., 2015). However, as the growth in computational power slows and physical models become increasingly complex, the bottlenecks of traditional NWP models are becoming more prominent. This is especially true in operational models, where trade-offs often need to be made between improving model resolution and increasing ensemble forecast members (Ben-Bouallegue et al., 2023), which substantially constrain the further development of NWP models.
Recently, researchers have begun exploring new and efficient weather forecasting paradigms, such as using machine learning (ML) algorithms including convolutional neural networks to predict future weather (Dueben and Bauer, 2018; Scher, 2018; Weyn et al., 2019; Rasp et al., 2020). Due to limitations in computational resources and ML algorithms at that time, these forecasts failed to surpass those from the Integrated Forecasting System of the European Centre for Medium-Range Weather Forecasts (ECMWF) (hereinafter referred to as ECMWF IFS). However, with ongoing advancements in ML algorithms, the Pangu-Weather (Pangu) model, which employs an advanced three-dimensional Earth-specific transformer algorithm (Bi et al., 2023), has achieved forecast scores that surpass those from the ECMWF IFS for the first time and operates over 10,000 times faster than traditional NWP models. Meanwhile, other innovative meteorological models such as GraphCast (Lam et al., 2023), FourCastNet (Pathak et al., 2022), Fengwu (Chen et al., 2023) and the Artificial Intelligence/Integrated Forecasting System (AIFS) (Lang et al., 2024) have also been introduced, leading a new wave of revolution in meteorology. Although models like the Pangu model currently achieve higher anomaly correlation coefficients for 500 hPa geopotential height than ECMWF IFS, they still face several limitations. These include a lack of physical interpretability (Bonavita et al., 2024), weaker intensity forecasts for tropical cyclones, and poor predictability for extreme weather events (Olivetti et al., 2024). Additionally, current ML models only provide a limited number of forecast variables, lacking comprehensive outputs of traditional NWP models, which restricts their application for general weather analysis.
Given the aforementioned limitations and considerations, there is an urgent need to integrate ML models with traditional physical models. But how can these two approaches be effectively combined? One common method is to use forecast fields from ML models as lateral boundary conditions to drive physical models, or to constrain physical model forecasts by using ML forecast fields through nudging techniques. Recently, Xu et al. (2024) were the first to propose using forecast fields from the Pangu model to drive the Weather Research and Forecasting (WRF) model, which significantly improved extreme precipitation forecasts in Beijing in 2023. Furthermore, Liu et al. (2024) utilized the Pangu model to drive the WRF model for tropical cyclone forecasts extending beyond two weeks, incorporating changes in sea surface temperature into the hybrid model, which further enhanced tropical cyclone forecasts. In this study, we employ the spectral nudging technique to integrate Pangu forecasts with WRF forecasts. Additionally, we investigate, for the first time, the role of data assimilation in the hybrid system, demonstrating that data assimilation can enhance typhoon forecasts even when constrained by Pangu forecasts.
The remainder of this paper is organized as follows. Section 2 describes the model and methods. The results are analyzed in section 3. Finally, section 4 gives the main conclusions and discussion.

\section{Model and Methods}
\subsection{Pangu-Weather model}
In this study, Pangu forecasts are implemented by using the ECMWF ai-models library (https://github.com/ecmwf-lab/ai-models). The Pangu model has a horizontal resolution of 0.25°×0.25° and provides upper-level forecast variables across 13 pressure levels (1,000 hPa, 925 hPa, 850 hPa, 700 hPa, 600 hPa, 500 hPa, 400 hPa, 300 hPa, 250 hPa, 200 hPa, 150 hPa, 100 hPa and 50 hPa), including geopotential height, temperature, specific humidity, and the u- and v-components of wind. Surface variables include surface temperature, mean sea level pressure, 2 m temperature, and 10 m u- and v-components of wind. In this study, the initial fields for Pangu are sourced from ERA5, with a forecast leading time of 5 days (120 hours) and an output interval of 6 hours. Pangu exhibits pronounced advantages in mid- to long-term forecasts of geopotential height or temperature fields compared with traditional NWP models. From Figs. 1a–1c, it is evident that the temperature field forecasts of Pangu align more closely with ERA5 compared with the Global Forecast System (GFS), particularly regarding the position and intensity of the warm core of Typhoon Doksuri (2023). The correlation coefficients between Pangu forecasts and ERA5 remain consistently above 0.99 within 120 hours, significantly outperforming those from the GFS (Fig. 1d). Additionally, the correlation coefficients for Pangu show a brief increase around every 24 hours, which is attributed to the model being trained with leading times of 1, 3, 6 and 24 hours, resulting in different iteration steps for each leading time.

\subsection{WRF configurations}
The operational Shanghai Typhoon Model (SHTM), which employs WRF version 3.8, is initialized daily at 0000 UTC (coordinated universal time, the same below) and 1200 UTC, providing forecasts for the subsequent 120 hours. The WRF model domain has 953×701 grid points with a horizontal resolution of 9 km, covering the region of 0.5°S–61.5°N, 58.3°E–172.7°E (black curve in Fig.1a). In this study, the model top is set at 50 hPa, consistent with the configuration of the Pangu model. The SHTM is initialized by using the GFS analysis and forecasts at a 0.25°×0.25° resolution, referred to as the GFS\_INIT experiment in this study. The model configuration includes the Thompson microphysics scheme (Thompson et al., 2004), the multi-scale Kain-Fritsch cumulus scheme (Budakoti et al., 2019), the general circulation model version of the rapid radiative transfer model for both longwave and shortwave (Mlawer et al., 1997), the unified Noah land-surface model (Ek et al., 2003), and the Yonsei University planetary boundary layer scheme (Hu et al., 2013), as detailed in Table 1.

\subsection{Spectral Nudging}

Nudging is a widely used downscaling technique in weather and climate research, generally categorized into grid nudging and spectral nudging (Mai et al., 2017). Grid nudging, being a stronger constraint, often limits high-resolution regional models from developing their own small-scale features, which can weaken the representation of typhoon intensity. Spectral nudging operates by decomposing environmental variables at specific wavelengths. However, introducing small-scale fields from global models into regional models may also hinder the intensification and formation of typhoons. Conversely, using a longer cutoff wavelength can reduce the impact of spectral nudging by excluding the large-scale fields crucial for determining typhoon tracks. Therefore, based on previous research (Gómez and Miguez-Macho, 2017; 2020), the cutoff wavelength is typically set to the Rossby radius (\textasciitilde1000 km), as this choice yields the optimal nudging effect. The nudging equation is expressed as follows (Moon et al., 2018):

\[
\frac{\partial \xi}{\partial t} = \mathcal{F}(\xi) + \beta \left( \xi^L_{\text{Pangu}} - \xi^L_{\text{WRF}} \right), \tag{1}
\]
where $\xi$ denotes the prognostic variable being nudged, and $\mathcal{F}$ represents the model operator. $\beta$ denotes the nudging coefficient and is set to the WRF default value of $0.0003 \, \text{s}^{-1}$. $\xi^L_{\text{Pangu}}$ and $\xi^L_{\text{WRF}}$ represent the large-scale components of the Pangu forecasts and the WRF forecasts, respectively. Variables used for nudging include the zonal and meridional wind components, temperature, and perturbation geopotential height (excluding specific humidity). To avoid disrupting typhoon dynamics near the surface, nudging is not applied to the 10 lowest sigma levels (approximately below 800 hPa).
\section{Results}

\subsection{Typhoon Predictions}
This study primarily compares the forecasts generated by the operational SHTM model using GFS as the initial field to drive the WRF model (i.e., GFS\_INIT) with those generated by using Pangu nudging to drive the WRF model (referred to as Pangu\_SP). Additionally, forecasts from the ECMWF IFS, recognized as the most advanced numerical model, are used as a reference as well. For Typhoon Doksuri, a total of four forecast experiments are conducted, with initial times set at 0000 UTC on July 21, 1200 UTC on July 21, 0000 UTC on July 22 and 1200 UTC on July 22, 2023. It can be found that in all four experiments, the typhoon tracks predicted by Pangu-SP outperform those predicted by the GFS\_INIT (Figs. 2a–2d). Notably, in the experiments initialized at 0000 UTC and 1200 UTC on July 21, 2023, the GFS\_INIT forecasts predict typhoon landfall over Taiwan, whereas the Pangu-SP forecasts predict the track through the Strait of Malacca, which aligns more closely with the observed track. Additionally, except for the experiment initialized at 1200 UTC on July 21, 2023, where Pangu\_SP slightly underperforms compared with the ECMWF IFS in terms of track prediction, Pangu-SP consistently outperforms the ECMWF IFS at other times. Over the 0–120 hour forecast period, the average track error of Pangu-SP is reduced by 20.3\% and 1.6\% compared with the GFS\_INIT and the ECMWF IFS, respectively. Since spectral nudging combines the large-scale advantages of Pangu with the small-scale and mesoscale strengths of WRF, Pangu\_SP shows a marked improvement in mean sea level pressure forecasting over ECMWF IFS (Figs. 2e–2h). In the experiments initialized at 0000 UTC and 1200 UTC on July 21, 2023, the intensity forecasts from Pangu\_SP are generally consistent with those from the GFS\_INIT (Figs. 2e and 2f). However, in the experiments initialized at 0000 UTC and 1200 UTC on July 22, 2023, the intensity predicted by Pangu\_SP is noticeably weaker compared with that predicted by the GFS\_INIT (Figs. 2g and 2h).

Furthermore, we compare the cloud field forecasts between Pangu\_SP and the GFS\_INIT. Due to the strong non-Gaussian nature of cloud field variables, such as cloud water and cloud ice, existing ML models have yet to achieve accurate cloud field forecasts. However, the ML-driven WRF model in this study enables the testing and diagnosis of multiple variables, which is critical for weather analysis. Figures 3a–3c illustrate the spatial distributions of observed brightness temperature from channel 12 (10.7 micron) of the Advanced Geostationary Radiation Imager (AGRI) onboard Fengyun-4B (FY-4B), as well as 72-hour cloudy simulations from the GFS\_INIT and Pangu\_SP with the initial time at 0000 UTC on July 22, 2023. The all-sky simulations of brightness temperature from FY-4B AGRI are generated by using the Community Radiative Transfer Model v2.3.0 after performing spatio-temporal interpolation of the WRF forecast fields to match the observation locations. These simulations require various input data, including cloud water content, effective radii for different hydrometeors (liquid water, ice, rain, snow and graupel), and vertical profiles of atmospheric temperature and water vapor mixing ratio. Additionally, surface variables such as surface temperature, 10 m wind and 2 m temperature, as well as satellite geometry variables, are also essential. The effective radii for liquid water, ice, rain, snow and graupel particles are set to 20 µm, 40 µm, 400 µm, 600 µm and 800 µm, respectively (Niu and Zou, 2024), with other input variables derived from the WRF forecasts. As shown in Fig. 3, the brightness temperature distribution from Pangu\_SP (Fig. 3c) is more consistent with observations (Fig. 3a), whereas the GFS\_INIT not only exhibits larger positional deviations but also underestimates brightness temperature in the typhoon core, indicating an overestimation of high clouds. Additionally, the distribution of brightness temperature below 240 K in Pangu\_SP aligns closely with observations, while GFS\_INIT overestimates the cloud heights of the deep convective clouds. In the 240 K–280 K range, both Pangu\_SP and the GFS\_INIT underestimate brightness temperature, with Pangu\_SP performing slightly better (Fig. 3d). This suggests that the models have limited capability in accurately capturing mid- and low-level clouds, highlighting the need for further investigation in future studies.

\subsection{Impact of Data Assimilation}
In this section, we explore whether data assimilation can further enhance the performance of the hybrid model under spectral nudging. The operational SHTM utilizes the Gridpoint Statistical Interpolation (GSI) data assimilation system. Currently, the SHTM employs a three-dimensional variational assimilation scheme and assimilates clear-sky data from the FY-4B AGRI water vapor channels in real time to improve the moisture distribution in the initial field (referred to as Pangu\_SPDA). Among the channels, the mid-level water vapor channel 10 (\textasciitilde6.95 micron) is highly correlated with the upper-level channel 9 (\textasciitilde6.25 micron) and the lower-level channel 11 (\textasciitilde7.42 micron), which leads to overlapping information and affects the assimilation effectiveness. Therefore, this study only assimilates water vapor channels 9 and 11. The observation error for these channels is set to 2.2 K, with a static background error covariance matrix. The bias correction is handled by using a variational bias correction scheme. Figures 4a–4d display the spatial distributions of the assimilated clear-sky observation-minus-background and observation-minus-analysis values from FY-4B AGRI water vapor channels 9 and 11 at 0000 UTC on July 22, 2023. It can be found that before assimilation, channels 9 and 11 exhibit prominent negative observation-minus-background (O-B) values. After assimilation, the observation-minus-analysis (O-A) values approximate a normal distribution with a mean of zero, indicating that the assimilation has effectively improved the analysis field. Additionally, the track errors of Pangu\_SPDA are consistent with those of Pangu\_SP, with slight improvements. However, Pangu\_SPDA substantially enhances the typhoon intensity forecasts compared with Pangu\_SP, making them largely consistent with the GFS\_INIT. Across the four different initial times, Pangu\_SPDA shows an average improvement of 2.2\% in intensity forecasting within the 0–120-hour period compared with Pangu\_SP. Consequently, Pangu\_SPDA achieves notably lower track errors compared with the GFS\_INIT, while its intensity forecasts remain consistent with those of the GFS\_INIT. This demonstrates that Pangu\_SPDA effectively combines the strengths of Pangu in typhoon forecasting with the advantages of WRF in typhoon intensity prediction.

\section{Conclusions and discussion}
This study utilizes spectral nudging to constrain the WRF forecast fields with the large-scale forecast fields from the Pangu model. Across four different initial times for Typhoon Doksuri, the Pangu\_SP experiment improves the track forecast by an average of 20.3\% within 120 hours compared with the GFS\_INIT experiment and by 1.6\% compared with the ECMWF IFS. For intensity forecasts, Pangu\_SP noticeably outperforms the ECMWF IFS, whereas it has larger errors and overall weaker forecasts compared with the GFS\_INIT. This study is the first to investigate whether data assimilation in the initial field can enhance the forecast performance of the ML-driven hybrid model. Additionally, it reveals that assimilating water vapor channels 9 and 11 of FY-4B AGRI substantially improves typhoon intensity forecasts, making them largely consistent with the observations. This indicates that data assimilation enhances the forecasting capabilities of the ML-driven hybrid model, particularly at the mesoscale.
Notably, this study uses a cutoff wavelength of 1000 km, corresponding to the Rossby radius. However, the suitability of this parameter for the hybrid model requires further investigation. Future research should also focus on enhancing the typhoon forecast performance of the hybrid model. This can be approached in two ways: by employing higher-performance models, such as the Artificial Intelligence/Integrated Forecasting System developed by the ECMWF, to nudge the WRF model, or by assimilating more localized data, such as from the Geostationary Interferometric Infrared Sounder onboard FY-4B, to improve the forecast performance. Furthermore, whether increasing the resolution of the hybrid model to 3 km or even 1 km enhances the accuracy of typhoon track and intensity forecasts also warrants further exploration.

\section*{Open Research}
FY-4B AGRI data were collected from the National Satellite Meteorological Center China Meteorological Administration (http://satellite.nsmc.org.cn/portalsite/default.aspx)

\acknowledgments
This research was supported by National Key Research and Development Program of China (Grants 2021YFC3000805).

\begin{figure}
    \centering
    \includegraphics[width=1\linewidth]{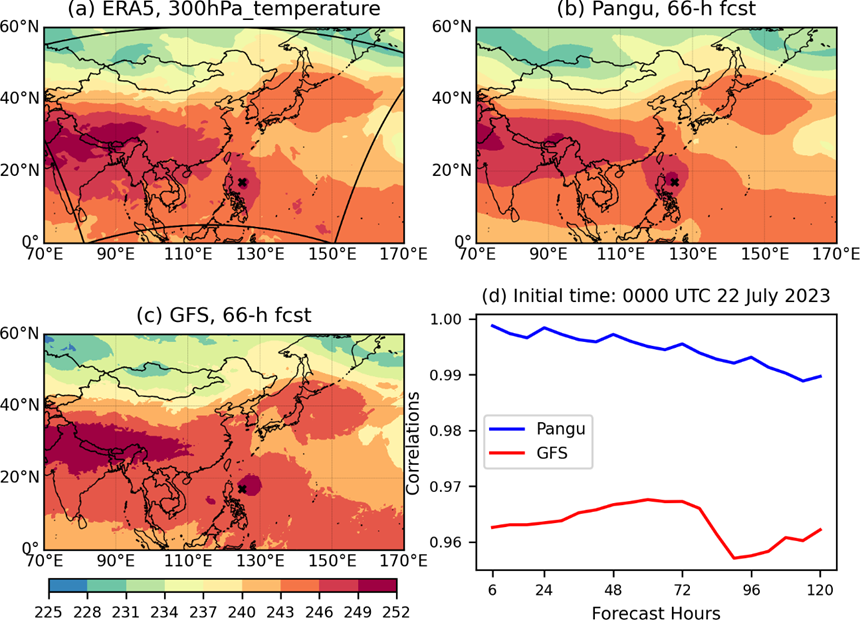}
    \caption{Spatial distributions of (a) the 300 hPa temperature field from ERA5, as well as 66-hour forecasts from (b) the Pangu-Weather (Pangu) model and (c) the Global Forecast System (GFS), with the initial time at 0000 UTC on July 22, 2023. (d) Correlations of the 300 hPa temperature field between Pangu and ERA5 (the blue line) and between GFS and ERA5 (the red line) versus forecast hours, with the initial time at 0000 UTC on July 22, 2023. The black box denotes the domain of the Weather Research and Forecasting (WRF) model, and black crosses represent the observed typhoon center.}
    \label{fig:enter-label}
\end{figure}

\begin{figure}
    \centering
    \includegraphics[width=1\linewidth]{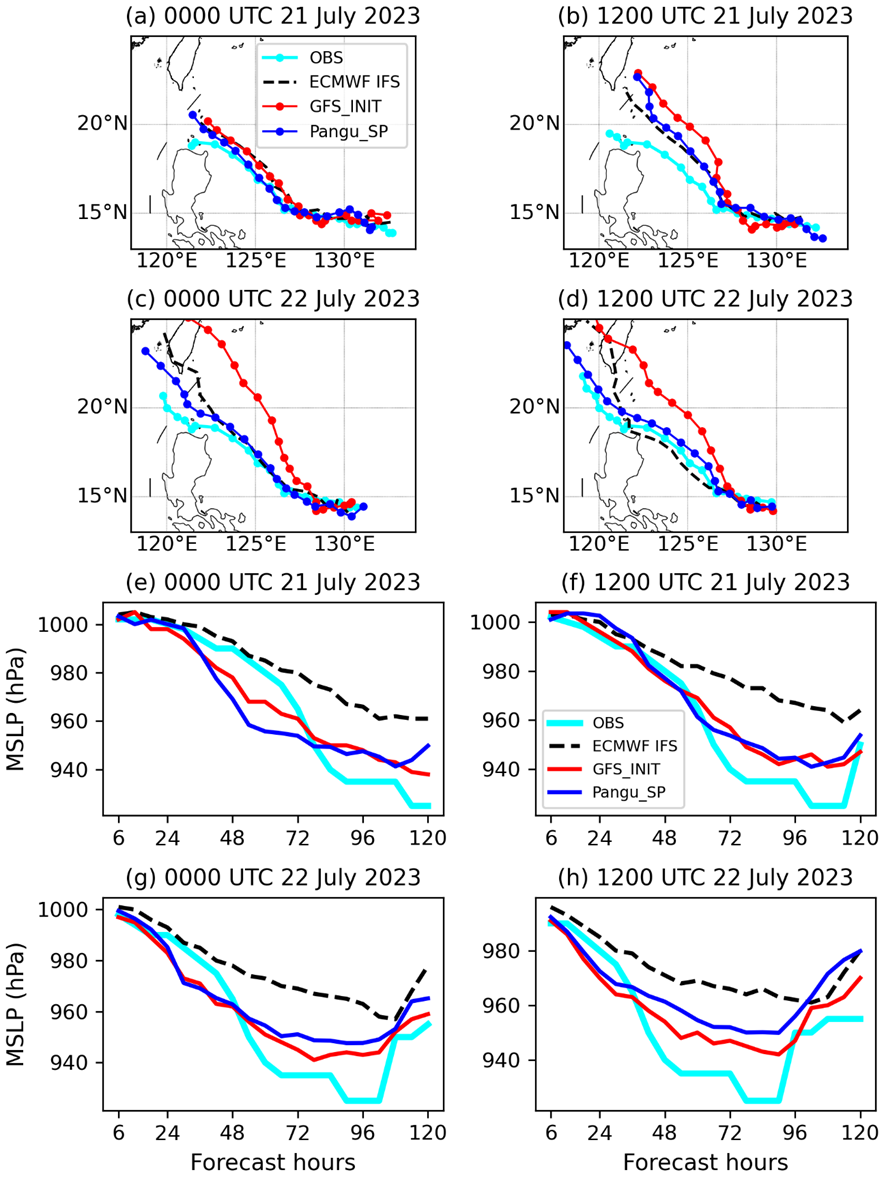}
    \caption{(a–d) Best tracks and (e–f) mean sea level pressure (cyan solid lines) for typhoon Doksuri (2023), along with 0–120 hour forecasts from the Integrated Forecasting System of the European Centre for Medium-Range Weather Forecasts (ECMWF IFS; black dashed lines), forecasts using the GFS as the initial field (GFS\_INIT; red solid lines), and forecasts generated by using Pangu nudging to drive the WRF model (Pangu\_SP; blue solid lines) at four different initial times.}
    \label{fig:enter-label}
\end{figure}

\begin{figure}
    \centering
    \includegraphics[width=1\linewidth]{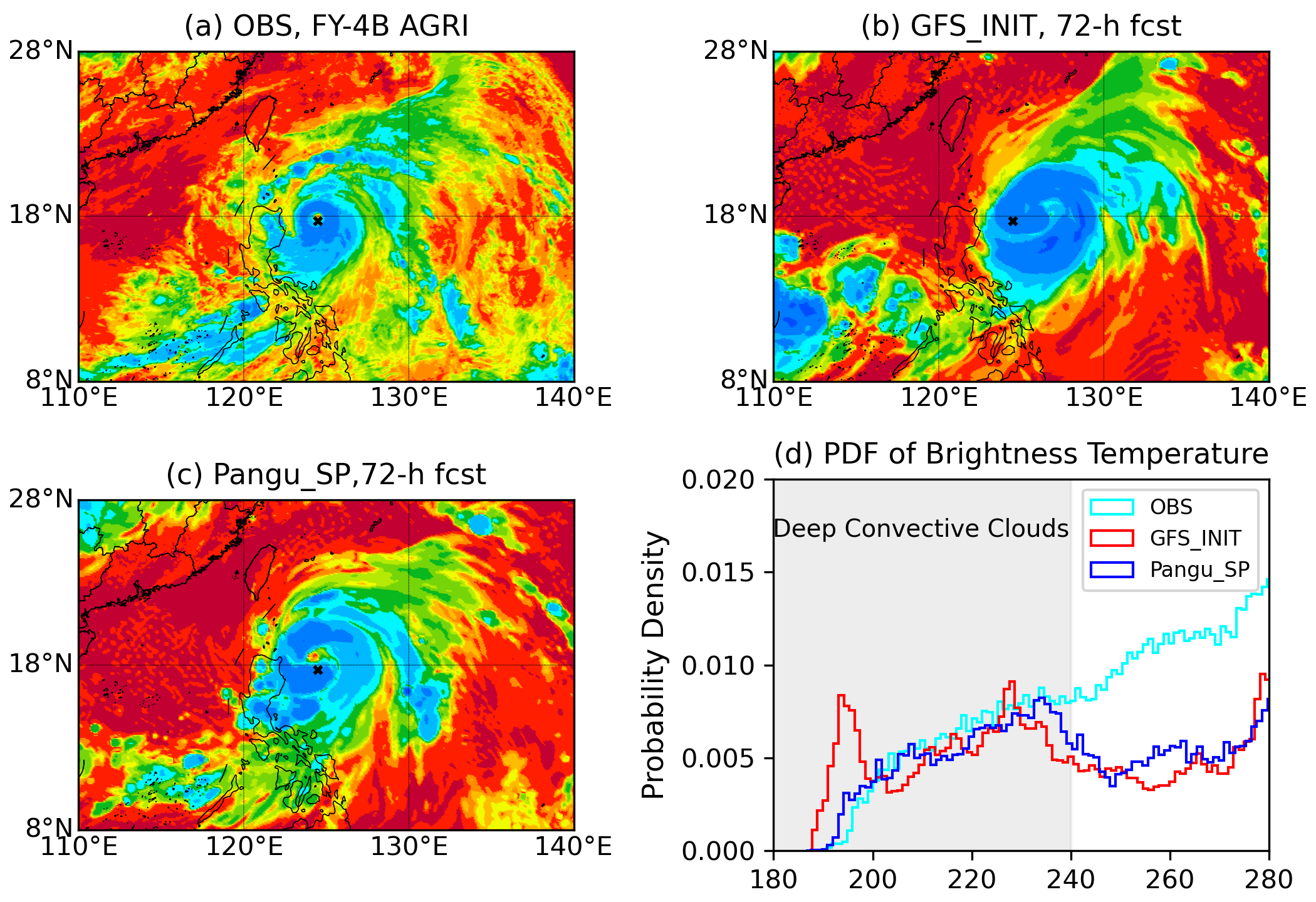}
    \caption{Spatial distributions of (a) observed brightness temperature (K) from the channel 12 (\textasciitilde10.7 micron) of the Advanced Geostationary Radiation Imager onboard Fengyun-4B, as well as 72-hour cloudy simulations from the (b) GFS\_INIT and (c) Pangu\_SP with the initial time at 0000 UTC on July 22, 2023. (d) Probability density of observed and simulated brightness temperature.}
    \label{fig:enter-label}
\end{figure}

\begin{figure}
    \centering
    \includegraphics[width=1\linewidth]{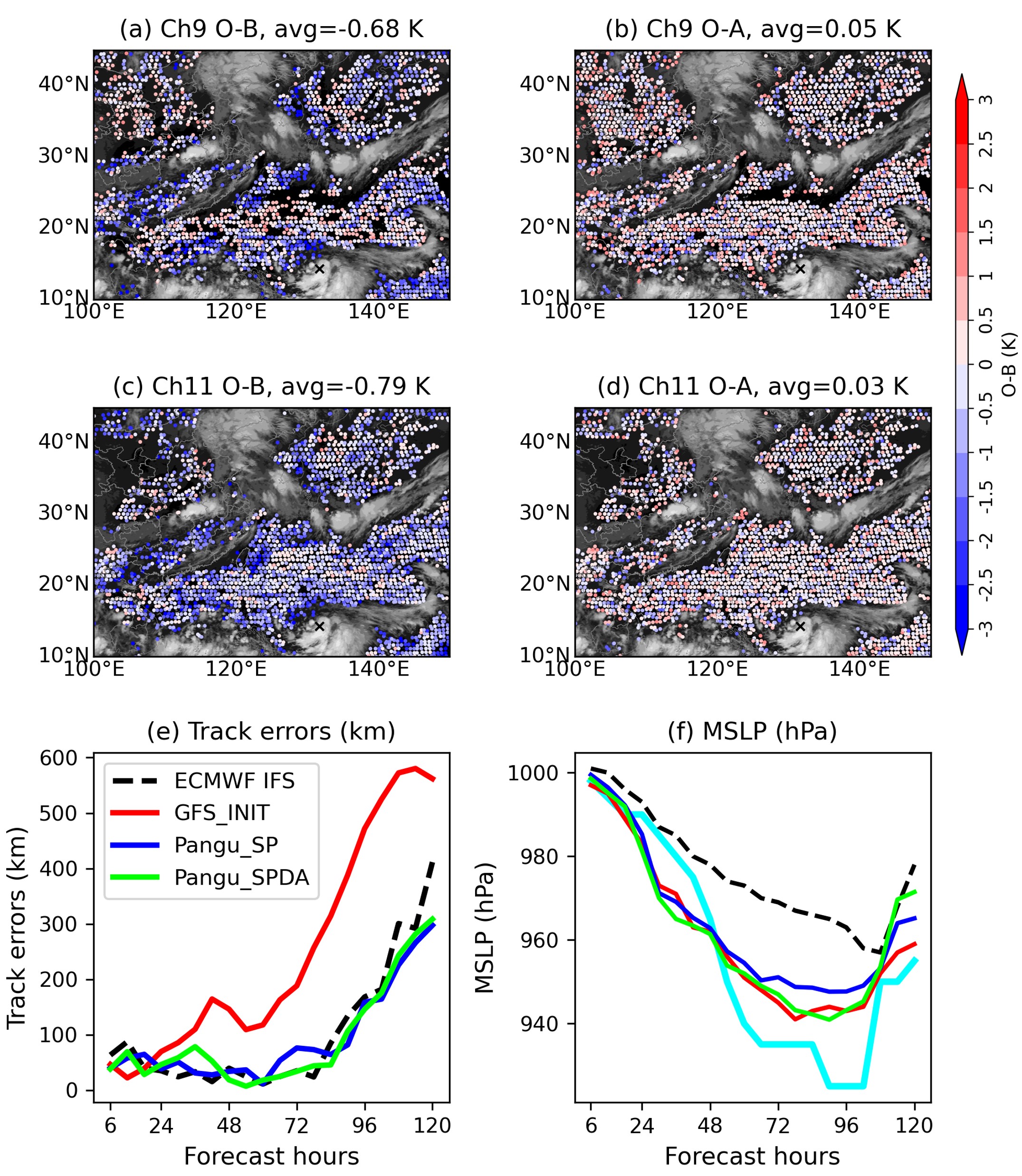}
    \caption{(a–d) Spatial distributions of the assimilated clear-sky observation-minus-background (O-B) and observation-minus-analysis (O-A) values from FY-4B AGRI water vapor channels 9 and 11 at 0000 UTC on July 22, 2023. (e) Track errors and (f) mean sea level pressure from the ECMWF IFS (black dashed lines), GFS\_INIT (red solid lines), Pangu\_SP (blue solid lines) and Pangu\_SPDA (green solid lines; i.e., the experiment of assimilating FY-4B AGRI water vapor channels based on the Pangu\_SP experiment), with the initial time at 0000 UTC on July 22, 2023.}
    \label{fig:enter-label}
\end{figure}

%
%


%
%
%
%
%

\end{document}